# Empirical facts characterizing banking crises: an analysis via binary time series[1]

Paolo Di Caro[1], Giuseppe Pernagallo[2], Antonino Damiano Rossello[3] and Benedetto Torrisi[4]

*1 Department of Finance, Italian Ministry of Economy and Finance; Rome, Italy. (email: paolo.dicaro@uniroma1.it)*
*2 Department of Economics and Business, University of Catania (**corresponding author**, e-mail: giuseppepernagallo@yahoo.it)*
*3 Department of Economics and Business, University of Catania (e-mail: rossello@unict.it)*
*4 Department of Economics and Business, University of Catania (e-mail: btorrisi@unict.it)*

**Abstract**

Various works have already showed that common shocks and cross-country financial linkages caused the banking systems of several countries to be highly interconnected with the result that during bad times, banking crises may arise simultaneously in different countries. Our aim is to provide further evidence on the topic using a dataset made by dichotomous banking crises time series for 66 countries from 1800 to 2014. Via the use of heatmap matrices we show that several countries exhibit pairwise correlation, which means that banking crises tend to occur in the same year. Clustering analysis suggests that developed countries (for the most European ones) are highly similar in terms of the path of events. An analysis of the events that followed the Great Depression and the Great Recession shows that after the crisis of 2008, banking crises tend to characterize countries tied by financial links whereas before 2008 contagion seems to affect countries in the same geographical area. Clustering analysis shows also that after financial liberalization crises affected countries with similar economic structures and growth. Further researches should enlighten the origin of these linkages investigating how the process of contagion eventually happens.

---

[1] Please do not quote without permission from the authors.

## 1. Introduction

The research question of this paper is straightforward: are banking crises in different countries associated? Various works have already showed that common shocks and cross-country financial linkages caused the banking systems of several countries to be highly interconnected (Bordo and Murshid, 2001; Neal and Weidenmier, 2003; Reinhart and Rogoff, 2013) with the result that during bad times, banking crises may arise simultaneously in different countries. Our aim is to provide further evidence on the topic using a big panel dataset. The period covered by our data goes from 1800 to 2014 for 66 countries; the series investigated are binary series classified with a 1 if a banking crisis occurred, 0 otherwise.

The methodology applied in this paper is intuitive: we first tested via runs test if randomness played any role in the sequence of events for each series, we then computed the pairwise correlation coefficients (for binary data) and tested their significance, presenting this huge number of coefficients via heatmap matrices. Via a chi-square test we found that there is no association between the intensity of correlation (moderate or strong) and the fact that the considered countries are in the same Continent or not. Finally, clustering analysis consented us to evidence group of nations similar in terms of the occurrence of banking crises assessing also how the two major financial crises and financial liberalization affected the considered countries.

The paper is structured as it follows. Section 2 briefly defines banking crises and what are the common causes provided in the literature. Section 3 shows the results of our analysis. The final section concludes.

## 2. Banking crises: definition and aetiology

The principal categories of risks faced by banks are credit risk (assets turn bad and ceasing to perform), liquidity risk (withdrawals exceed the available funds), and interest rate risk (fluctuations in the interest rate can reduce the value of bonds held by the bank). These risks, if not correctly managed by banks, can be source of financial instability.

There is not a unanimous definition of banking crisis (Chaudron and Haan, 2014). As pointed out by the *Word Bank*, a (systemic) *banking crisis* can be defined as the situation of many banks in a country that experience solvency or liquidity problems simultaneously. When this situation occurs, the financial sector of that country is characterized by several defaults of financial institutions, and a large part of the institutions face the impossibility or the difficulty of repaying contracts on time. Calomiris (2010, p. 4) defines banking crisis as "*panics or waves of bank failures*". Another definition concerns the side of capital, so a banking crisis identifies a situation of a large depletion of the banking system capital (Caprio and Klingebiel, 1996; Laeven and Valencia, 2008; Reinhart and Rogoff, 2009; Laeven, 2011).

The causes of a banking crisis can be different (Laeven, 2011). An important role is played by *bank runs* and *bank panic*. If depositors of a bank suspect the bank to fail, they can rush to withdraw their deposits; if too many depositors follow the same behaviour, the bank can be forced to liquidate its assets until failure (Diamond and Dybvig, 1983). When many bank runs occur at the same time and concern different banks, we talk of *bank panic*, which results are catastrophic and cause the failure of many financial institutions. Bank runs are closely related to the irrational side of depositors: the fear that the other depositors can withdrawn their funds causing the bank to fail, triggers the run even if the bank is financially stable (*self-fulfilling prophecies*). In order to prevent this fear, nowadays the international normative provides several protections for depositors such as higher reserve requirements for banks and deposit guarantees.

Others think that banking crises are the natural consequence of the procyclicality of the financial system (Gorton, 1988), so during upturn of the economic cycle, credit grows consistently whereas during the downturn there is a collapse in credit. These theories are linked to economic fundamentals and can be thought as the opposite of theories that justify banking crisis as the consequence of bubbles. Asset price bubbles exist when "*the market price of an asset exceeds its price determined by fundamental factors by a significant amount for a prolonged period*" (Evanoff, Kaufman and Malliaris, 2012), a situation that cannot be adequately explained by traditional theories and clearly in contrast with the assumption of informational efficiency.

Frauds can also be source of banking crises. Fraudulent behaviour can cause the failure of several banks, which in turns triggers a banking crisis. Some remarkable examples were the crises of Venezuela in 1994 and in the Dominican Republic in 2003 (Laeven, 2011).

Finally, given the fact that many banks over the world are interconnected, financial contagion is another source of crises: the failure of large intermediaries in one country can cause the failures of other intermediaries in other countries. Financial contagion can be thought as a sort of "domino effect" (Freixas, Laeven and Peydrò, 2015), in the sense that the failure of a bank can cause failures of other intermediaries. The magnitude of this effect depends on the overall macroeconomic and financial environment and can be detrimental for the banking system since it increases the probability of default of an intermediary and the fragility of the financial system (Freixas, Laeven and Peydrò, 2015; for a definition of financial fragility see Lagunoff and Schreft, 2001). Freixas, Laeven and Peydrò (2015) stated that contagion can start via seven channels. Among these, our work provides evidence regarding the channel of cross-country contagion. The recent financial crisis has been challenging to economists because of its multiple-country nature showing a systemic crisis in a single "small" country is completely different from a worldwide systemic crisis. Due to financial environments highly interconnected, a crisis can boost in a country and spreads quickly in other countries. The evidence supporting the relationship between banks over the word is well documented. Reinhart and Rogoff (2013, p. 4561) pointed out that in the late-2000s crisis, the origin of the widespread financial instability was due to "*common shocks (the bursting of the global housing bubble) and cross-country linkages (for example, because many countries bought US subprime mortgage debt)*". Other authors (Bordo and Murshid, 2001; Neal and Weidenmier, 2003) showed the existence of cross-country correlations in banking crises also during the period 1880–1913. Kalemli-Ozcan, Papaioannou, and Perri (2013), using a sample of 20 developed countries, showed that countries more interconnected with the United States experienced more synchronized output cycles with the United States during the financial crisis of 2008. In this paper we want to extent the evidence in favour of cross-country linkages in banking crises assessing whether a relationship between crises in different countries exists and whether this eventual relationship can be considered statistically significant.

The novelty of our work is clear: we provide evidence in favour of the fact that banking crises occur simultaneously using a big panel dataset, including countries with different economic and social structures. Furthermore, we show how these crises tend to interest clusters of countries that present economic or political ties, investigating also the differences occurred with respect to the two major financial crises (1929 and 2008) and after the financial liberalization in 1990s.

## 3. Data and results

As pointed out by Chaudron and Haan (2014), recent researches on banking crises generally use as sources for dating banking crises the works of Caprio et al. (2005), Reinhart and Rogoff (2009) and Laeven and Valencia (2008, 2013). These databases consider a (systemic) banking crisis when exceptional events or policy measures occur.

The dataset used in this study is available on the website of *Harvard Business School*. These data (collected over many years by Carmen Reinhart, Ken Rogoff, Christoph Trebesch, and Vincent Reinhart) include banking crises for several countries from 1800 to 2014. The dataset is made by binary time series taking value of 1 when a banking crisis occurred, 0 otherwise. The 66 countries used in this study are reported in Table 1. The United Kingdom and United States are the countries with the highest number of banking crises (both 33), whereas the countries with the lowest number of crises are Singapore and Mauritius with only 1 banking crisis in more than 200 years. The most nefarious year in terms of number of crises was the 1995, with 26 banking crises over the world; the first advanced-economy banking crisis in the considered period occurred in 1802 in France, whereas the first emerging markets interested by the phenomenon were India, China and Peru (Reinhart and Rogoff, 2013).

The first hypothesis to verify is whether the sequence of events for each country follows a random path. The binary series is made of a sequence of 0s and 1s; in such cases the nonparametric *runs test* is a common methodology used to test the randomness of a series. A run can be defined as an unbroken sequence of similar events or like objects (Bradley, 1960). For example, in the series AABABBBAA there are five runs: one run of A's of length 1, two runs of A's of length 2, one run of B's of length 1 and one run of B's of length 3. If the outputs A and B are randomly distributed, we should not recognize any pattern in the series. We tested if the occurrence of banking crises in a country follows a random pattern (null hypothesis) or a determined pattern, i.e. the sequence of events is not casual (alternative hypothesis). Performing this test before we compute the correlation matrix is important to exclude any role of randomness in the sequence of events, which may cause spurious correlation.

Table 2 shows the results of the test reporting both an exact p-value and a p-value obtained via approximation. For almost every country we can reject the null hypothesis of randomness of the series concluding that the occurrence of banking crises is not casually distributed, however, the null hypothesis cannot be rejected for Guatemala, whereas there is contrasting evidence for the Dominican Republic, Honduras, Mauritius and Singapore (it should be noted that the low number of crises of these countries may affected the result of the test).

***Table 1.*** *Countries used for the empirical analysis.*

| Country | Label | Country | Label |
|---|---|---|---|
| Algeria | DZA | Kenya | KEN |
| Angola | AGO | Korea | KOR |
| Argentina | ARG | Malaysia | MYS |
| Australia | AUS | Mauritius | MUS |
| Austria | AUT | Mexico | MEX |
| Belgium | BEL | Morocco | MAR |
| Bolivia | BOL | Myanmar | MMR |
| Brazil | BRA | Netherlands | NLD |
| Canada | CAN | New Zealand | NZL |
| Central African Republic | CAF | Nicaragua | NIC |
| Chile | CHL | Nigeria | NGA |
| China | CHN | Norway | NOR |
| Colombia | COL | Panama | PAN |
| Costa Rica | CRI | Paraguay | PRY |
| Cote D'Ivoire | CIV | Peru | PER |
| Denmark | DNK | Philippines | PHL |
| Dominican Republic | DOM | Poland | POL |
| Ecuador | ECU | Portugal | PRT |
| Egypt | EGY | Russia | RUS |
| El Salvador | SLV | Singapore | SGP |
| Finland | FIN | Spain | ESP |
| France | FRA | Sri Lanka | LKA |
| Germany | DEU | Sweden | SWE |
| Ghana | GHA | Taiwan | TWN |
| Greece | GRC | Thailand | THA |
| Guatemala | GTM | Tunisia | TUN |
| Honduras | HND | Turkey | TUR |
| Hungary | HUN | United Kingdom | GBR |
| Iceland | ISL | United States | USA |
| India | IND | Uruguay | URY |
| Indonesia | IDN | Venezuela | VEN |
| Italy | ITA | Zambia | ZMB |
| Japan | JPN | Zimbabwe | ZWE |

*Source: our elaboration.*

**Table 2.** Runs test for the binary series of banking crises.

| Country | Exact p-value | P-value via approximation | Country | Exact p-value | P-value via approximation |
|---|---|---|---|---|---|
| DZA | 0.0016 | <0.0001 | KEN | <0.0001 | 0.0000 |
| AGO | <0.0001 | 0.0000 | KOR | <0.0001 | 0.0000 |
| ARG | <0.0001 | <0.0001 | MYS | <0.0001 | 0.0000 |
| AUS | <0.0001 | <0.0001 | MUS | 0.0186 | 0.9228 |
| AUT | 0.0023 | <0.0001 | MEX | <0.0001 | <0.0001 |
| BEL | <0.0001 | <0.0001 | MAR | 0.0187 | <0.0001 |
| BOL | 0.0002 | <0.0001 | MMR | <0.0001 | 0.0000 |
| BRA | 0.0002 | <0.0001 | NLD | <0.0001 | 0.0000 |
| CAN | 0.0014 | <0.0001 | NZL | <0.0001 | 0.0000 |
| CAF | <0.0001 | 0.0000 | NIC | <0.0001 | 0.0000 |
| CHL | 0.0002 | <0.0001 | NGA | <0.0001 | 0.0000 |
| CHN | <0.0001 | <0.0001 | NOR | <0.0001 | 0.0000 |
| COL | <0.0001 | 0.0000 | PAN | 0.0187 | <0.0001 |
| CRI | <0.0001 | 0.0000 | PRY | <0.0001 | 0.0000 |
| CIV | <0.0001 | 0.0000 | PER | <0.0001 | 0.0000 |
| DNK | <0.0001 | <0.0001 | PHL | <0.0001 | 0.0000 |
| DOM | 0.1106 | 0.0002 | POL | <0.0001 | 0.0000 |
| ECU | <0.0001 | 0.0000 | PRT | <0.0001 | <0.0001 |
| EGY | <0.0001 | 0.0000 | RUS | <0.0001 | <0.0001 |
| SLV | 0.0016 | <0.0001 | SGP | 0.0186 | 0.9228 |
| FIN | 0.0023 | <0.0001 | ESP | <0.0001 | 0.0000 |
| FRA | <0.0001 | <0.0001 | LKA | <0.0001 | 0.0000 |
| DEU | <0.0001 | <0.0001 | SWE | <0.0001 | 0.0000 |
| GHA | <0.0001 | 0.0000 | TWN | 0.0023 | <0.0001 |
| GRC | <0.0001 | 0.0000 | THA | <0.0001 | 0.0000 |
| GTM | 0.1101 | 0.8204 | TUN | <0.0001 | 0.0000 |
| HND | 0.0558 | <0.0001 | TUR | 0.0002 | <0.0001 |
| HUN | <0.0001 | 0.0000 | GBR | <0.0001 | <0.0001 |
| ISL | <0.0001 | 0.0000 | USA | <0.0001 | <0.0001 |
| IND | <0.0001 | 0.0000 | URY | <0.0001 | 0.0000 |
| IDN | <0.0001 | 0.0000 | VEN | <0.0001 | 0.0000 |
| ITA | <0.0001 | <0.0001 | ZMB | <0.0001 | 0.0000 |
| JPN | <0.0001 | 0.0000 | ZWE | <0.0001 | 0.0000 |

*Source: our elaboration.*

In order to measure the relationship between banking crises of two different countries, we used tetrachoric correlation coefficients. Since the correlation matrix in this case is very big (a 66 x 66 matrix), we represented the data via a heatmap matrix computed over the whole period[2]. Figure 1 shows the results of the correlation between the series, whereas Figure 2 is the matrix of the statistically significant coefficients associated to the matrix in Figure 1.

The principal features of the analysis can be synthesized as it follows:

- the agreement between the series is for the most part positive; only few coefficients are negative and the most of these are not statistically significant.
- Overall, we see that there is high positive correlation between banking crises in different countries. This means that if a coefficient between two countries, say A and B, is positive (and statistically significant), then banking crises in A are associated with banking crises in B, and vice versa (for example, the coefficient for DZA and AUS, or KOR and COL).
- As it emerges from Figure 2, the relationship measured via tetrachoric correlation is for a large part of the coefficients significative.
- The strength of correlation is independent from the geographical location of the considered couple of countries. This is evident from the matrix, since countries located in different Continents shows high and significative correlation (for example, the correlation between NIC and EGY, two distant countries, is high and significative at 1%), but we tested statistically this assumption via a chi-square test, considering in rows the degree of correlation (moderate[3] or strong) and in columns if the considered couple of countries were located in the same Continent or not[4]. The resulting p-value is about 0.3, which means that we cannot reject the hypothesis of independence.

---

[2] Considering a shorter period would be counter-productive: first, because for inference, the more the data, the better the estimates; second, because computing such matrices is computationally expansive, therefore we opt for the most representative matrix; third, because banking crises are "rare" events, excluding precious observations from the analysis could affect the magnitude and the significance of the coefficients; finally, because more matrixes would occupy too much space in the paper without adding different point of views.
[3] A moderate correlation coefficient assumes values (in absolute terms) in the range [0,0.7[.
[4] This is a possible criterion of spatial proximity, simple but effective to consider the fact that countries in the same geographical area tend to exhibit similar financial and normative structures.

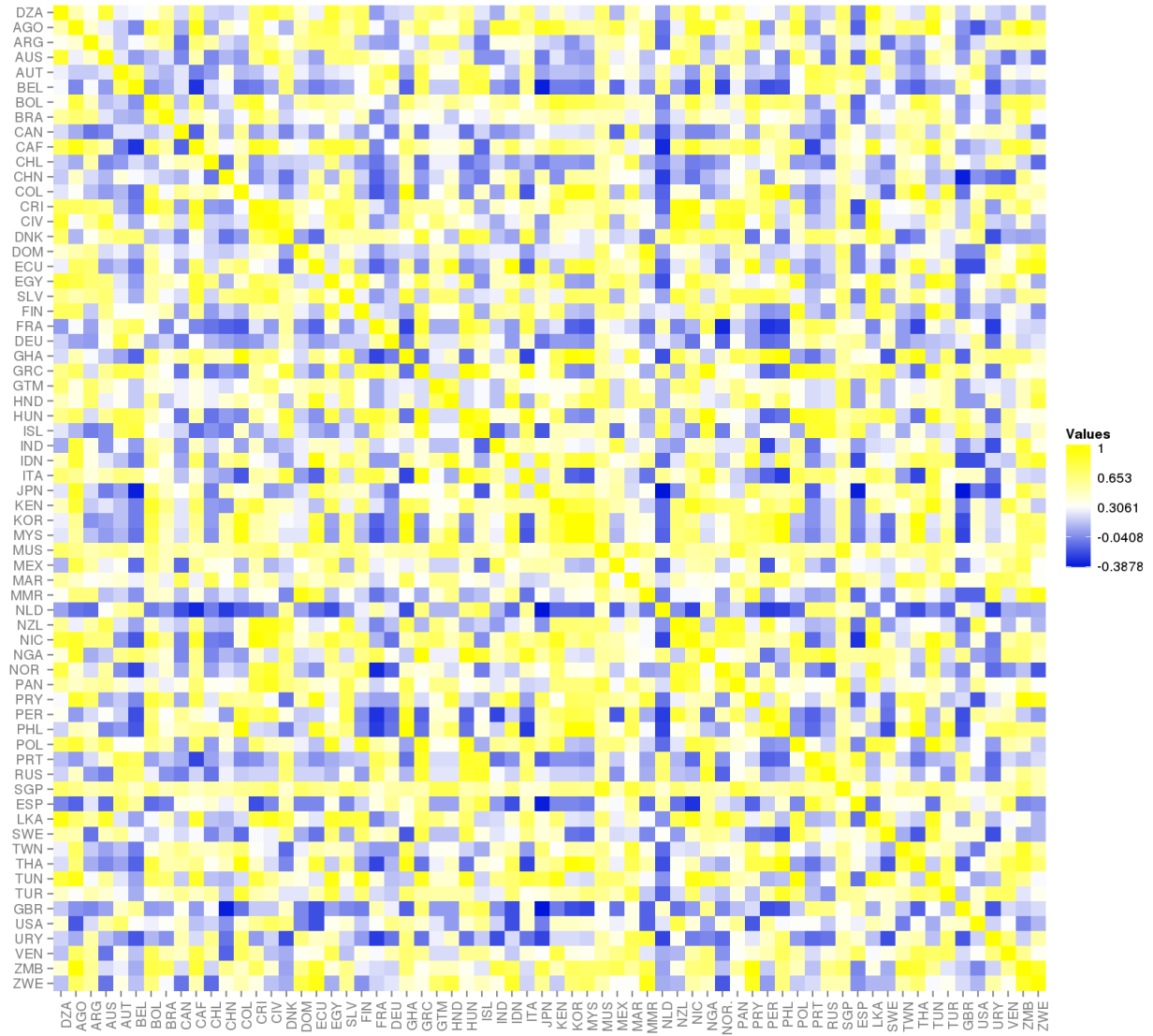

*Figure 1.* Heatmap matrix for pairwise tetrachoric correlation coefficients between dichotomous time series for the whole time period. Source: our elaboration.

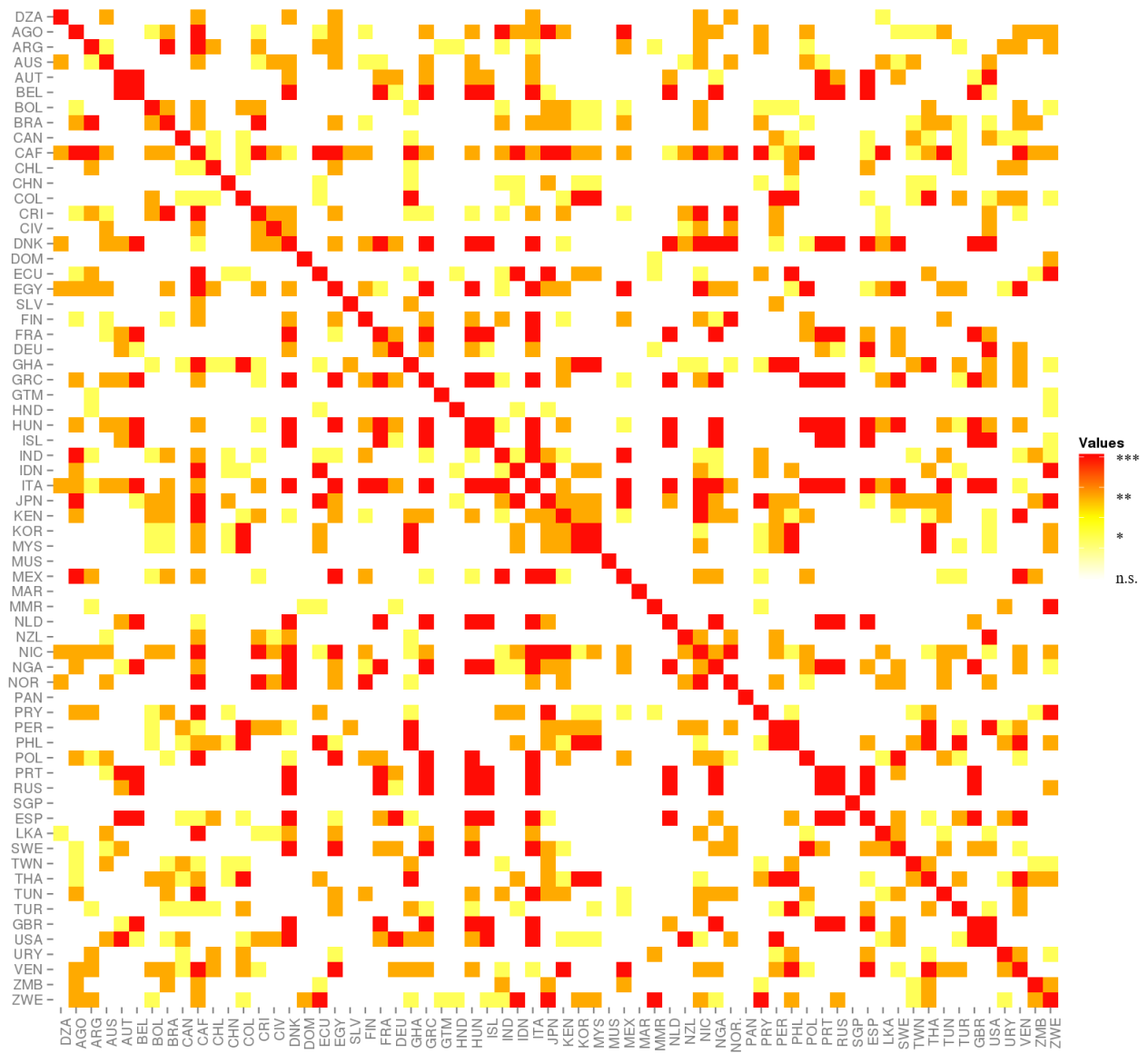

*Figure 2.* Matrix of the statistically significant correlation coefficients. * indicates a 10% significance, ** indicates a 5% significance, *** indicates a 1% significance, n.s. stands for "non-significant". The main diagonal is indicated in red for visualisation purpose. Source: our elaboration.

Finally, using clustering procedures we can visualise what groups of countries have been characterized by common events. Figure 3 shows the dendrogram for the 66 analysed countries over the whole period, using the Jaccard index as dissimilarity criterion and average linkage as aggregation method. From the dendrogram we can make interesting observations: we can see that Hungary and Greece have been characterized by the same path of events since their dissimilarity is zero; Greece is also closely related to Nigeria, whereas other strong similarities concern Polonia and Tunisia, Costa Rica and Nicaragua, Korea and Malaysia, Colombia and Philippines; Guatemala and Honduras form an isolated cluster and it should be noted that they are adjacent countries, while Singapore banking crises are not pooled to any other series; the light blue cluster is formed, except for Nigeria, by developed countries linked by political or economical relationships, furthermore, the majority of these countries are European; the cut-off threshold is very high, therefore almost all the groups are significative for the analysis.

It is interesting to see what clusters emerge if we consider the two major financial crises in 1929 and in 2008. For the purpose, we divided the sample into two sub-sample: the first one covers banking crises from 1929 to 2006 and takes into account the effects of the Great Depression; the second one covers banking crises from 2006 to 2014 and takes into account the effects of the Great Recession. Figure 4 shows the dendrogram for the first sub-sample whereas Figure 5 shows the dendrogram for the second one. Some cluster is like the situation described by the dendrogram for the entire period (for example Greece-Hungary, Singapore or Dominican Republic-Mauritius-Myanmar), but the situation considering the impact only of the Great Depression is very different and the main characteristics can be synthesized as follows. First, we note that Austria, Germany, Portugal and France form now a cluster separated by the USA; second, the blue cluster shows that there is a strong similarity among countries located in the same geographical are. It is the case, for example, of Norway-Denmark, Italy-Egypt or Angola-Nigeria. These two facts prove that before the Great Recession, banking crises were "contagious", but principally via a geographical criterion. After the Great Recession (Figure 5) it emerges a clear characteristic: banking crises now characterize countries tied by financial links. The dendrogram can be divided in two part: the first part, on the left, shows big cluster with no dissimilarity (the series is very short therefore several countries showed the same occurrence of events); the second part, on the right, shows a group formed for the most by European countries strictly related to USA.

A final distinction investigated in this paper regards the degree of financial liberalization. Since developing countries experienced an important series of reform about financial liberalization in the 1990s (World Bank, 2005), we divided the series into two sub-series: one considers the years from 1800 to 1990 and the second one the years after 1990 to assess the impact of financial liberalization. Figure 6 shows the dendrogram for the period before the principal reforms and Figure 7 the period after. Several important differences emerge from an analysis of the two dendrograms. Interestingly, after financial liberalization there is the formation of two big cluster (Figure 7) whereas before the series of reforms that interested the 1990s the situation was more mixed (Figure 6). After financial liberalization (Figure 7), there is a brown cluster that gathers, approximately, the major part of the developed countries (such as USA, France, Germany, UK or Italy etc.) whereas the major part emerging/developing countries (such as India, Indonesia, Taiwan or Mexico) are gathered by the grey cluster.

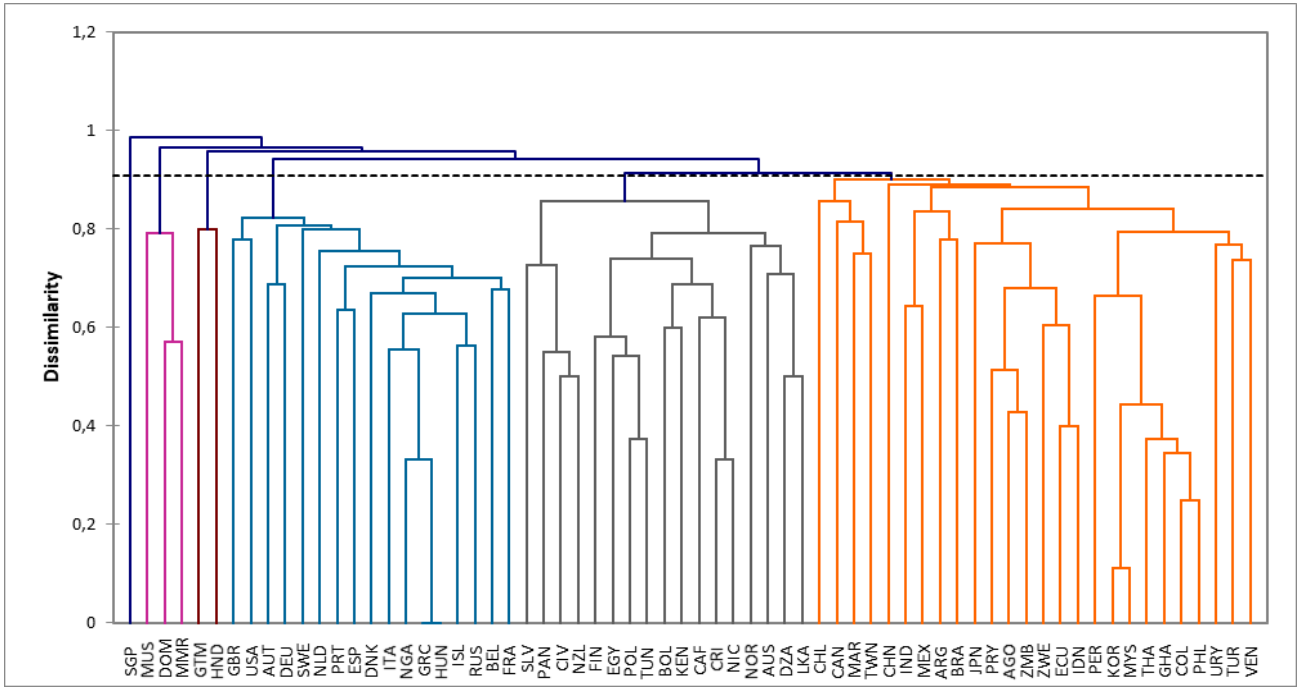

*Figure 3.* Dendrogram for the 66 banking crises series from 1800 to 2016. Measure of dissimilarity: Jaccard index. Aggregation method: average linkage. The dotted line is the cut-off threshold. Source: our elaboration.

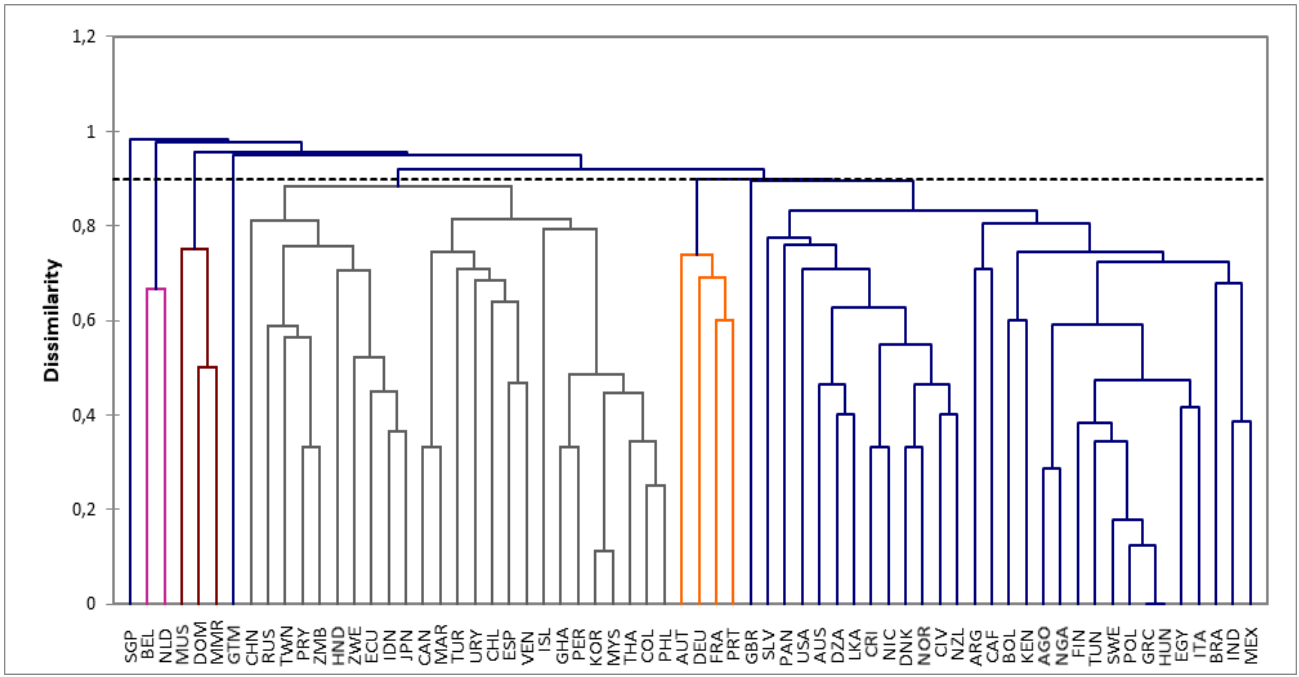

*Figure 4.* Dendrogram for the 66 banking crises series from 1929 to 2006. Measure of dissimilarity: Jaccard index. Aggregation method: average linkage. The dotted line is the cut-off threshold. Source: our elaboration.

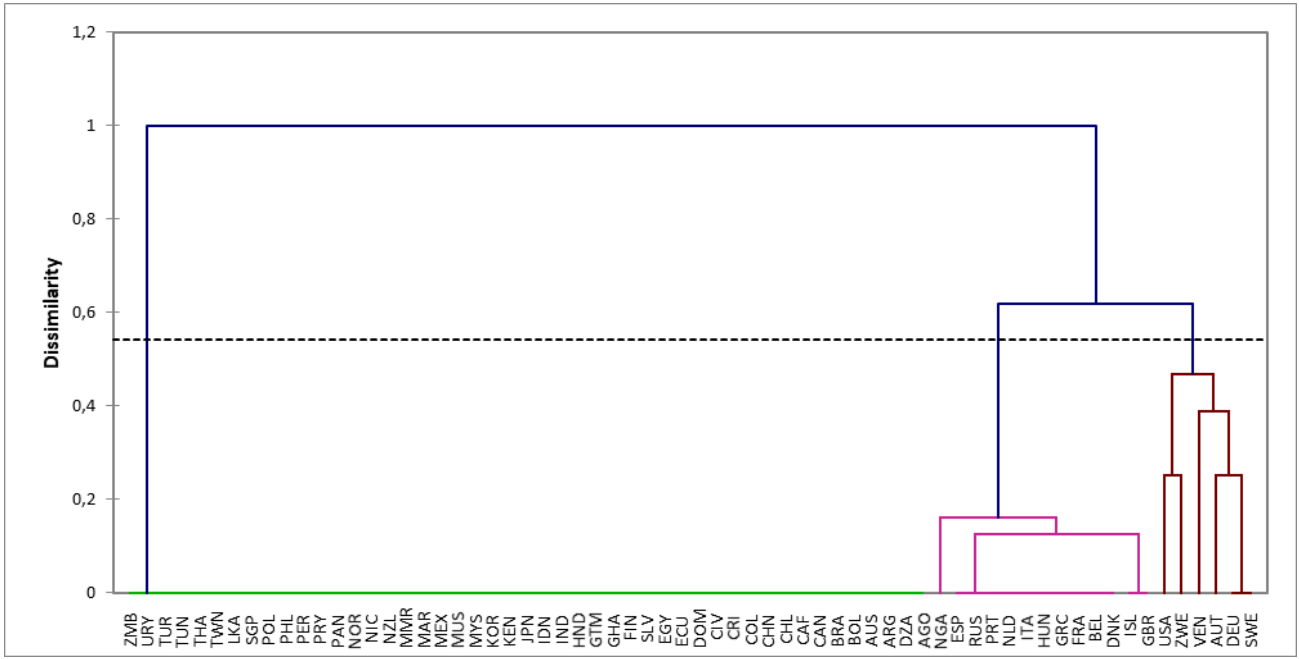

*Figure 5.* Dendrogram for the 66 banking crises series from 2007 to 2014. Measure of dissimilarity: Jaccard index. Aggregation method: average linkage. The dotted line is the cut-off threshold. Source: our elaboration.

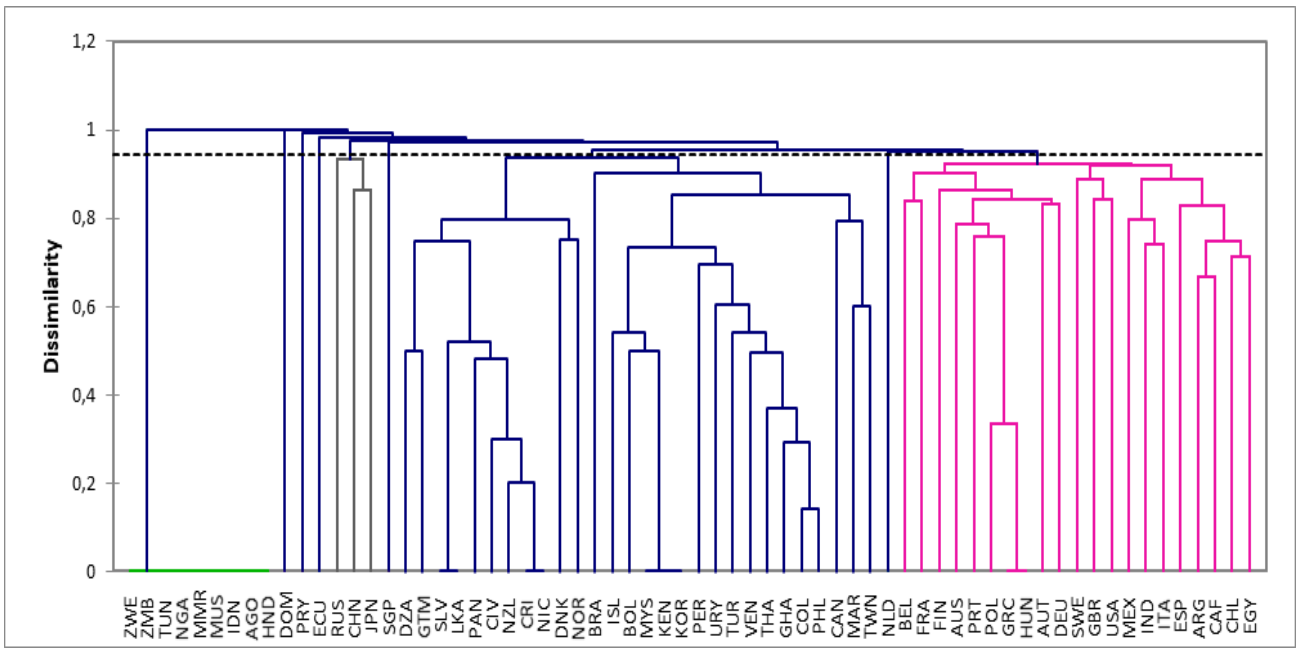

*Figure 6.* Dendrogram for the 66 banking crises series from 1800 to 1990. Measure of dissimilarity: Jaccard index. Aggregation method: average linkage. The dotted line is the cut-off threshold. Source: our elaboration.

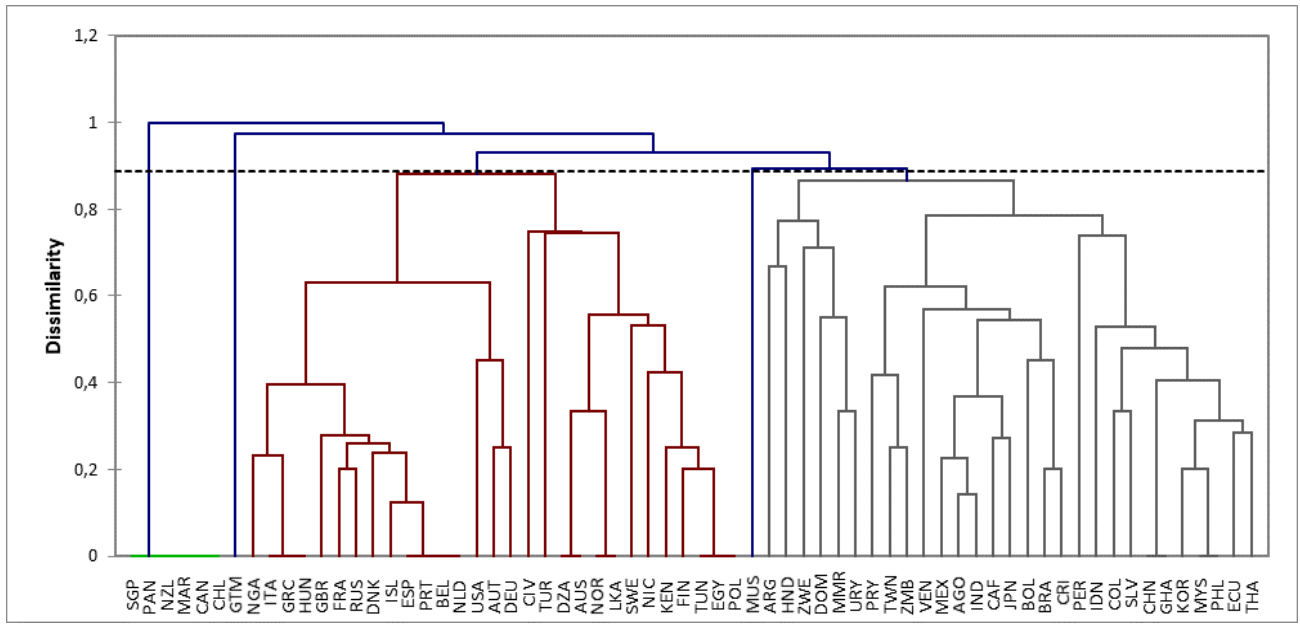

*Figure 7*

*Dendrogram for the 66 banking crises series from 1991 to 2014. Measure of dissimilarity: Jaccard index. Aggregation method: average linkage. The dotted line is the cut-off threshold. Source: our elaboration.*

## 4. Conclusions

In this paper we provided further evidence in favour of cross-country correlation between banking crises. First, via runs tests we showed that the occurrence of banking crises does not follow a random path (except for Guatemala). The result, although obvious, is important to exclude the role of causality in explaining a correlation between banking crises in different countries. Second, we computed tetrachoric correlation for 66 countries suggesting that crises in different countries are associated. This result is particularly important because enlightens the fact that banking crises in different places may have the same determinants and that financial contagion is a concrete danger in a global economy. Our analysis seems reliable because we studied the phenomenon over more than 200 years, we tested properly the significance of the coefficients, via runs test we excluded any role of randomness in the crisis series and via the chi-square test we found that the intensity of correlation is not dependent from the fact that two countries are located in the same geographical area (Continent). Via the use of clustering analysis we found out that banking crises that interested developed countries (especially in Europe) tend to be closely linked as they form a clear cluster. This points out that at least for certain economic or political areas, the causes of the crises may be the same. Clustering analysis also enlightened the fact that after the Great Recession banking crises characterized countries tied by financial links (for the major part developed countries) whereas after financial liberalization clusters were formed based on a development criterion.

The role of regulation is crucial for the issue. For example, Gluzmann and Guzman (2017) showed that liberalization reforms (concerning capital accounts, securities markets, interest rates, removal of credit controls, barriers to entry, and reduction of state ownership in the banking sector) in emerging economies were directly associated with a higher frequency of banking crises. Freixas, Laeven and Peydrò (2015) emphasized the role of macro-prudential policies in preventing the occurrence of financial crisis. These measures become indispensably to reduce the likelihood of banking crises because financial liberalization increased cross-border banking activities whereas financial globalization weakened the effectiveness of domestic policies, even though other studies found out that a greater banking sector globalization and financial liberalization diminish the likelihood of a banking crisis (Shehzad and Haan, 2009; Ghosh, 2016). Future researches should deepen our knowledge about the common causes of banking crises and how the process of contagion spreads from one country to another.